\documentclass[superscriptaddress,aps,prl,twocolumn,showpacs,preprintnumbers,amsmath,amssymb,floatfix,groupedaddress]{revtex4}

\usepackage{graphicx}
\usepackage{dcolumn}
\usepackage{bm}

\begin{document}

\title{Alkaline-Earth-Metal Atoms as Few-Qubit Quantum Registers}

\author{A. V. Gorshkov}
\affiliation{Physics Department, Harvard University, Cambridge, Massachusetts 02138, USA}
\author{A. M. Rey}
\affiliation{JILA, National Institute of Standards and Technology and University of Colorado, Boulder, CO 80309-0440, USA}
\author{A. J. Daley}
\affiliation{\mbox{Institute for Theoretical Physics, University of Innsbruck, A-6020 Innsbruck, Austria
and Institute for Quantum Optics} and Quantum Information of the Austrian Academy of Sciences, A-6020 Innsbruck, Austria}
\author{M. M. Boyd}
\author{J. Ye}
\affiliation{JILA, National Institute of Standards and Technology and University of Colorado, Boulder, CO 80309-0440, USA}
\author{P. Zoller}
\affiliation{\mbox{Institute for Theoretical Physics, University of Innsbruck, A-6020 Innsbruck, Austria
and Institute for Quantum Optics} and Quantum Information of the Austrian Academy of Sciences, A-6020 Innsbruck, Austria}
\author{M. D. Lukin}
\affiliation{Physics Department, Harvard University, Cambridge, Massachusetts 02138, USA}

\date{\today}

\begin{abstract}

We propose and  analyze a novel approach to quantum information processing, in which multiple qubits can be encoded and manipulated using electronic and nuclear degrees of freedom 
associated with individual alkaline-earth atoms trapped in an optical lattice. Specifically, we describe how the qubits within each register can be individually manipulated and measured with sub-wavelength optical resolution. We also show how  such few-qubit registers can be  coupled to each other in optical superlattices via conditional tunneling to form a scalable quantum network. Finally, potential applications to quantum computation and precision measurements are discussed. 

\end{abstract} 




\maketitle

Stringent requirements on the implementation of scalable quantum information systems can be significantly relaxed if the entire system can be subdivided into smaller quantum registers,  in which several qubits can be stored for a long time and local quantum operations can be carried out with a very high fidelity \cite{dur03sorensen03moehring07saffman, childress05, jiang}. In such a case, scalable quantum networks can be built even if the non-local coupling between registers has either low fidelity or is probabilistic \cite{dur03sorensen03moehring07saffman, jiang, childress05}. Local operations can be achieved with the highest fidelity if the entire register is encoded into a single atom or molecule. While
quantum registers based on individual solid-state impurities are already being explored \cite{dutt07jiang08stutt}, typical qubits encoded into hyperfine  \cite{chaudhury07} or Rydberg \cite{ahn00} states of isolated atoms cannot be easily used as a few-qubit register. 
More specifically, the Hilbert space associated with such systems cannot be represented as a direct product of several sub-systems, such that e.g.\ one of the qubits can be measured without affecting the others.

In this Letter, we show that individual 
alkaline-earth atoms can be used for
the robust implementation of quantum registers, with one (electronic) qubit encoded in a long-lived optical transition and several additional qubits encoded in the nuclear spin. Following the 
proposal of Ref.\ \cite{childress05}, 
we use the electronic qubit as the communication qubit \cite{jiang, dutt07jiang08stutt} for detecting and coupling the registers. 
In particular, we show that the full $(2 I + 1)$-dimensional space describing a spin-$I$ nucleus  can be preserved during electronic-qubit detection.
This step uses off-resonant detection proposed in Ref.\ \cite{childress05} and extends the 
proposal of Ref.\ \cite{reichenbach07} beyond one nuclear qubit manipulation to much larger registers  ($I$ can be as high as $9/2$, as in $^{87}$Sr). 
We also show how to manipulate and measure individual registers in an optical lattice with subwavelength resolution.
While entangling gates between alkaline-earth atoms have been studied in the context of nuclear qubits alone \cite{hayes07,daley} and electronic qubits alone \cite{stock08}, we propose a new scheme that makes use of 
both degrees of freedom. 
Our gate creates entangled states between electronic qubits using conditional resonant tunneling and an interaction blockade \cite{folling07, cheinet08}. 

\begin{figure}[b]
\includegraphics[scale = 0.83]{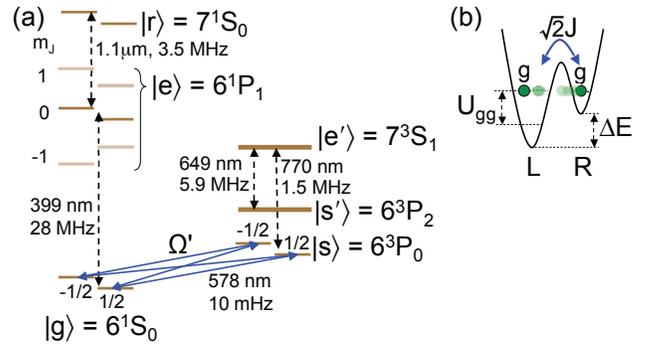}
\caption{(color online) (a) On the example of ${}^{171}$Yb ($I = 1/2$) in the Paschen-Back regime for state ${}^1P_1$, 
 relevant alkaline-earth-like level structure. (b) Interregister gate based on conditional resonant tunneling.  
 \label{fig1}}
\end{figure}

\textit{The register.}---Fig.\ \ref{fig1}(a) shows, as an example, the relevant alkaline-earth-like structure of ${}^{171}$Yb ($I = 1/2$). We want to 
arbitrarily manipulate the $2 (2 I + 1)$-dimensional Hilbert space consisting of states in the manifolds $|g\rangle = {}^1S_0$ and $|s\rangle = {}^3P_0$ ("g" for ground and "s" for stable). Using a differential g-factor \cite{boyd07}, 
one can optically excite  all 6 I + 1
individual $|g\rangle-|s\rangle$ transitions in the presence of a magnetic field $B$ \cite{ye08} [see Fig.\ \ref{fig1}(a)]. Thus, any 
unitary operation on an individual register can be achieved \cite{brennen05,speednote}. 

To make single-register manipulation spatially selective, one can envision various strategies. The conceptually simplest strategy would adiabatically expand the lattice for the time of the manipulation, which can be done without changing the wavelength of the light using holographic techniques or angled beams \cite{peil03}. Alternatively, we can make a temporary Raman transfer of a given pair of Zeeman levels of $|s\rangle$ up to $|s'\rangle = {}^3P_2$  via $|e'\rangle = {}^3S_1$  [Fig.\ \ref{fig1}(a)] \cite{daley}. One way to achieve spatial selectivity in this Raman transfer is to perform adiabatic passage, in which the beam on one of the transitions vanishes (and, thus, prohibits the transfer) at the location of one atom \cite{cho07}. Subsequent manipulation of the two chosen Zeeman levels of $|s\rangle$ will then be automatically spatially selective. Any other pair of states within the register can 
 be 
  manipulated by swapping it in advance with the first one. Alternatively, one can use the $|s\rangle-|e'\rangle-|s'\rangle$ Lambda system to achieve 
  spatial selectivity even 
 without state transfer by using dark-state-based selectivity \cite{gorshkov08}. 

 \textit{Interregister Gate.}---We now 
assume that atoms are prepared in a Mott insulator state \cite{greiner02} in an optical lattice (or a fully polarized band insulator, which may be easier to load), with one atom per site.
We 
isolate two adjacent atoms (left and right) 
using a superlattice \cite{folling07, porto}.  We now show how to generate a two-qubit phase gate between the electronic qubits of these atoms (i.e.\ $|g,g\rangle \rightarrow - |g,g\rangle$) 
in a regime where the 
tunneling $J$ is much smaller than the 
onsite interaction energy. 
As shown in Fig.\ \ref{fig1}(b), 
we 
bias
the right well relative to the left well by a value $\Delta E$ equal to the interaction energy $U_{gg}$ between two $|g\rangle$-atoms. If we had bosons with $I = 0$, then after time $\tau \sim 1/J$ the 
state $|g,g\rangle$ would pick up a minus sign due to resonant tunneling of the right atom to the left well and back. 
We now demonstrate how this gate works for fermions with arbitrary $I$. We consider the two-well single-band Hubbard Hamiltonian \cite{lightshiftnote}
\begin{eqnarray}
\hat H_h &=&  - J  \sum_{\alpha, m}  (\hat c^\dagger_{L \alpha m}\hat c_{R \alpha m} + \textrm{h.c.}) + \Delta E  \sum_{\alpha,m} \hat n_{R \alpha m} \nonumber \\
&& - \sum_{i,\alpha,m}  \mu_N B m g_{\alpha} \hat n_{i \alpha m}  + \!\!\!\!  \sum_{i, \alpha, m < m'} \!\!\!\!  U_{\alpha \alpha} \hat n_{i \alpha m} \hat n_{i \alpha m'} \nonumber \\
&& + V \sum_{i} \hat n_{i g} \hat n_{i s} + V_\textrm{ex}\!\! \sum_{i, m, m'} \!\! \hat c^\dagger_{i g m} \hat c^\dagger_{i s m'} \hat c_{i g m'} \hat c_{i s m}. 
\end{eqnarray}
Here $i = L, R$ labels sites; $\alpha = g, s$; $m, m' = -I, \dots, I$; $\hat n_{i \alpha} = \sum_m \hat n_{i \alpha m}$; $\hat n_{i \alpha m} = \hat c^\dagger_{i \alpha m} \hat c_{i \alpha m}$, where $\hat c^\dagger_{i \alpha m}$ creates an atom in state $\alpha m$ on site $i$. The tunneling rate $J$ and the bias $\Delta E$ are assumed for notational simplicity to be state-independent. $g_\alpha$ is the $g$-factor of state $\alpha$. $V = (U^+_{sg} + U^-_{sg})/2$ and $V_\textrm{ex} = (U^+_{s g} -  U^-_{sg})/2$ describe 
the "direct" and "exchange" interactions \cite{roth66}. $U^x_{\alpha \beta}  = (4 \pi \hbar^2 a^x_{\alpha \beta}/M) \int d^3 r |\phi_\alpha (\mathbf r)|^2 |\phi_\beta (\mathbf r)|^2$ \cite{ho98}, where $M$ is atomic mass, $a_{\alpha \beta}^x = a_{gg}, a_{ss}, a_{sg}^+, a_{sg}^-$ are the four s-wave scattering lengths, 
 $\phi_\alpha$ are the Wannier orbitals. $a^-_{sg}$ 
 corresponds to the antisymmetric electronic state $|gs\rangle-|sg\rangle$ (implying a symmetric nuclear state), while $a_{gg}$, $a_{ss}$, and $a^+_{sg}$ correspond to the three symmetric electronic states (implying antisymmetric nuclear states). 
Since $|g\rangle$ and $|s\rangle$ have $J = 0$ and since 
hyperfine mixing of $|s\rangle$ with other states is small \cite{boyd07}, we take $a_{\alpha \beta}^x$ to be independent of nuclear spin, which is consistent with experiments \cite{boyd07b, ye08}. 
We note that the optical energy of $|s\rangle$ is absent in our rotating frame.


In our scheme, provided $U_{gg}$ differs from other $U$'s, the interaction blockade \cite{cheinet08} will prevent two atoms from being on the same site unless they are both in state 
$|g\rangle$, so we can 
ignore all but the $U_{gg}$ interaction terms. In this case, 
the 
Zeeman Hamiltonian 
can 
 be rotated out. 
The first step of the gate is to increase the bias $\Delta E$ from $0$ 
to $U_{gg}$ for time $\tau = \pi/(\sqrt{2} J)$, and then set it back to $0$. Defining $|g,g\rangle |m_2,m_1\rangle = \hat c^\dagger_{L g m_2} \hat c^\dagger_{R g m_1} |0\rangle$, this gives a $2 \pi$ pulse between $I (2 I + 1)$ 
states $|g,g\rangle (|m_2, m_1\rangle - |m_1, m_2\rangle)$ ($m_1 < m_2$) and $\hat c^\dagger_{L g m_1} \hat c^\dagger_{L g m_2} |0\rangle$, so that the former pick up a factor $-1$. 
The $I (2 I + 1)$ states that pick up the 
factor $-1$ are precisely all the $|g,g\rangle$ states with an antisymmetric nuclear state since two $|g\rangle$ atoms in a symmetric nuclear state cannot sit on one site. 
To make \textit{all} $(2 I + 1)^2$ $|g,g\rangle$ states 
pick up the 
factor $-1$, we require two 
more steps. In the second step, we apply a phase $-1$ on site $R$ on all $|g,m\rangle$ with $m > 0$, repeat the bias, and repeat the phase. In the final step, we swap $|g,m\rangle$ and $|g,-m\rangle$ on site $R$, repeat the first two steps, 
and repeat the swap. This results in 
$|g,g\rangle \rightarrow -|g,g\rangle$ independent of the nuclear spin, i.e.\ a two-qubit phase gate on the two electronic qubits. All atom pairs in the superlattice that experience only the four biases are unaffected. Thus, together with spatially selective single-atom manipulation, this gate gives universal manipulation of the full lattice of quantum registers. The gate error due to 
virtual tunneling is $\sim (J/U)^2$, where $J/U$ is the smallest relevant ratio of tunneling to interaction energy or to a difference of interaction energies. This error can be reduced if $|g\rangle$ and $|s\rangle$ lattices are independent \cite{daley}. 
Other errors in this gate are analogous to those studied in Ref.\ \cite{cheinet08}.

We now point out some advantages of this gate.  The gate is essentially achieved by conditioning the resonant tunneling on the internal state of the atoms rather than on the number of atoms in the wells \cite{folling07, cheinet08} or on the vibrational levels the atoms occupy \cite{strauch08}. Being resonant, the gate is faster ($\tau \sim 1/J$) than superexchange gates ($\tau \sim U/J^2$) \cite{trotzky08}. 
At the same time, by conditioning the tunneling on the internal state, we avoid having two $|s\rangle$-atoms in one well \cite{stock08}, which may be subject to collisional losses.
A key property of the gate is that it couples the electronic (communication) qubits without affecting the nuclear qubits. 
At the same time, a remarkable feature of our gate is that 
it would not have worked without the use of the nuclear degree of freedom, because two $|g\rangle$ atoms would not be able to sit on the same site in that case. This is in a sense the reverse of Ref.\ \cite{hayes07}, where a gate on nuclear spins relies on the underlying electronic interactions. 
Finally, our gate can be easily extended to bosons. In particular, 
a single bias interval would suffice for
bosons with two internal states $|g\rangle$ and $|s\rangle$ that have different interactions $U_{gg}$, $U_{ss}$, and $U_{sg}$ (e.g.\ if $|g\rangle$ and $|s\rangle$ experience different potentials). 


\textit{Electronic-qubit detection.}---We now demonstrate the essential ability of our register to preserve all nuclear qubits during the fluorescence detection of the electronic qubit. 
The key ingredients will be off-resonant excitation \cite{childress05} and/or a strong magnetic field \cite{reichenbach07}.
The detection is made by cycling the $|g\rangle - (|e\rangle = {}^1P_1)$ transition ("e" for excited). 
To yield an error $p < 0.01$ after scattering $N \sim 100$ photons,
the decay rate from $|e\rangle$ to $|g\rangle$ should exceed 
other decay rates from $|e\rangle$ by 
$> 10^4$, which is typically satisfied \cite{lellouch87loftus02xu03, branchingnote}. 
We can thus restrict ourselves to a $4 (2 I + 1)$-dimensional space describing the $|g\rangle -  |e\rangle$ transition: $|g\rangle |m_I\rangle$ ($J = 0$) and $|e,m_J\rangle |m_I\rangle$ ($J = 1$). 
The Hamiltonian is then ($\hbar = 1$) \cite{lightshiftnote} 
\begin{eqnarray}
\hat H = A  \mathbf{\hat I} \cdot  \mathbf{\hat J} + Q \frac{3 (\mathbf{\hat I} \cdot  \mathbf{\hat J})^2 + 3/2 \mathbf{\hat I} \cdot  \mathbf{\hat J} - K}{2 I J (2 I - 1) (2 J -1)} + g_J \mu_B \hat J_z B  \nonumber \\
 - g_I \mu_N \hat I_z B - \Omega (|g\rangle \langle e,0| + \textrm{h.c.}) - \Delta \sum_m |e, m\rangle \langle e, m|.  \label{ham}  
\end{eqnarray}
Here $K = I (I+1) J (J+1)$; $A$ and $Q$ ($Q = 0$ for $I = 1/2$) are the magnetic dipole and electric quadrupole hyperfine 
constants, respectively; 
$g_J$ and $g_I = g_g$ are the relevant 
$g$ factors; $\Omega$ and $\Delta$ are the Rabi frequency and the detuning of the 
$\pi$-polarized probe light. 
Using three Lindblad operators $\hat L_m = \sqrt{\Gamma} |g\rangle \langle e,m|$, 
the master equation is
\begin{equation}
\dot \rho = - i [\hat H,\rho] - \frac{1}{2} \sum_m (\hat L^\dagger_m \hat L_m \rho + \rho \hat L_m^\dagger \hat L_m - 2 \hat L_m \rho \hat L_m^\dagger).
 \label{master}
\end{equation}

\begin{figure}[b]
\includegraphics[scale = 0.99]{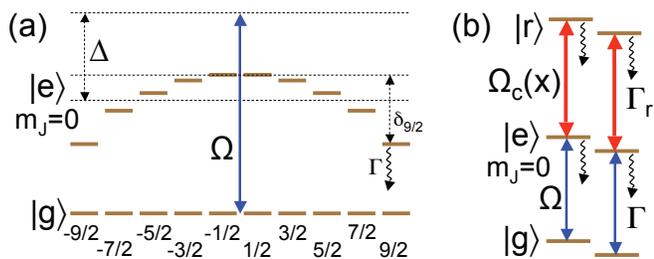}
\caption{(color online) Nuclear-spin-preserving electronic-qubit detection. (a) $^{87}$Sr ($I \!=\! 9/2$)
 in the Paschen-Back regime
(the shift $- g_I \mu_N  m_I B$ not shown). (b) On the example of ${}^{171}$Yb  ($I \!=\! 1/2$), 
dark-state-based spatial selectivity \cite{gorshkov08}.
 \label{fig2}}
\end{figure}

Two approaches  to preserve nuclear coherence during fluorescence are possible. In the first one, a strong magnetic field  ($g_J \mu_B B \gg A, Q$) decouples $\mathbf{\hat I}$ and $\mathbf{\hat J}$ in the Paschen-Back regime  \cite{reichenbach07}. In the second one, a large detuning $\Delta \gg A, Q$ does the decoupling by means of the interference of Raman transitions via all excited states with a given $I_z + J_z$ \cite{childress05}. Unless $Q \ll \Gamma$, the first approach fails because the frequencies of  transitions $|g\rangle|m_I\rangle - |e,0\rangle |m_I\rangle$ differ by $\delta_{m_I} = 3 Q m_I^2$ [see Fig.\ \ref{fig2}(a)]. However, when $Q \ll \Gamma$, the first approach may be preferable as it allows for much faster detection than the second (off-resonant) approach. While the two approaches can work separately, their combination is sometimes advantageous and allows for the following simple estimate of the nuclear decoherence due to off-resonant excitation. We assume $\Delta \gg Q, \Omega, \Gamma$ and a magnetic field large enough to decouple $\mathbf{\hat I}$ and $\mathbf{\hat J}$ (arbitrary $B$ can be analyzed similarly). The number of photons scattered during time $\tau$ is then $N \approx \Gamma \tau \Omega^2/\Delta^2$. Furthermore, for any two $m_I = m_1, m_2$, the four coherences $|g/e,0\rangle|m_1\rangle - |g/e,0\rangle |m_2\rangle$ form a closed system. Adiabatically eliminating the three coherences except for the ground one, the latter is found to decay with rate $\Gamma_{12} \approx (\delta_{m_1} - \delta_{m_2})^2 \Omega^2 \Gamma/(2 \Delta^4)$, yielding an error $p \sim \Gamma_{12} \tau \sim N (Q/\Delta)^2$. Thus, to scatter $N = 100$ photons and obtain $p < 0.01$, we need $\Delta \gtrsim 100 Q$. 

To verify low $p$ numerically, 
we use $^{171}$Yb ($I = 1/2$) 
\cite{hoyt05}, $^{87}$Sr ($I = 9/2$) \cite{ye08}, and $^{43}$Ca ($I = 7/2$), for which 
$(\Gamma, A, Q)/(2 \pi \textrm{ MHz}) = (28, -213, 0)$  
\cite{reichenbach07},
$(30.2, -3.4, 39)$ 
\cite{reichenbach07},
and $(35, -15.5, - 3.5)$ \cite{nortershauser98}, respectively. Although less widely used, $^{43}$Ca has the advantageous combination of small $Q$ and large $I$. 
We prepare the atom in some state in the manifold $|g\rangle$, turn $\Omega$ on and off abruptly for a time $\tau$, and then wait for all the population to decay down to $|g\rangle$, which transforms Eq.\ (\ref{master}) into a 
superoperator $\mathcal{\hat E}$ acting on density matrices describing $|g\rangle$. Ideally, $\mathcal{\hat E}$ 
describes a unitary transformation $\hat U$ that maps $|g\rangle |m\rangle \rightarrow \exp(i \phi_m) |g\rangle |m\rangle$ with 
$\phi_{-I} = 0$ and $\phi_m$ ($m > -I$) given by the phase of the diagonal elements of $\mathcal{\hat E}$ corresponding to the density matrix element $\rho_{m,-I}$. So $p \equiv 1 - \bar F$, where $\bar F$ is the average gate fidelity of $\mathcal{\hat E}$ with respect to $\hat U$ \cite{nielsen02}: 
\begin{equation}
\bar{F}(\mathcal{\hat E},\hat U) \equiv \int d \psi \langle \psi| \hat U^\dagger \mathcal{\hat E}(|\psi\rangle \langle \psi|) \hat U| \psi\rangle.
\end{equation}

We fix $N = 100$ and begin by considering the first approach (large $B$ and $\Delta = 0$). In Yb, $B = 2$T and $\Omega/2 \pi = 30$ MHz ($\tau = 1.3 \mu$s) give 
$p \approx 0.01$. Since Yb has $I = 1/2$ (hence $Q =0$), as $B \rightarrow \infty$, $p \rightarrow 0$: for example, $B = 10$ T and $\Omega/2 \pi = 200$ MHz ($\tau = 1.1 \mu$s) give 
$p \approx 10^{-4}$. In Ca, small $Q/\Gamma$ ($\approx 0.1$) also allows one to obtain high fidelity on resonance:  
$B = 1$T and $\Omega/2 \pi = 200$ MHz ($\tau = 0.9 \mu$s) give 
 $p \approx 0.002$.  Since $Q$ is finite here, 
 increasing $B$ further does not 
 reduce $p$ to zero.  Finally, in Sr, resonant scattering 
 gives $p \gtrsim 0.1$ due to the large $Q$. Turning now to the second approach ($B = 0$ and large $\Delta$), for Yb, Ca, and Sr, $(\Delta,\Omega)/(2 \pi \textrm{ GHz}) = (15, 0.2)$, $(6,0.07)$, and $(3,0.04)$, respectively, give $\tau \sim 3$ ms and $p \approx 0.01$.  An increase of $\Omega$ (to reduce $\tau$) leads unfortunately to larger $p$ at least partly due to the loss of adiabaticity in the evolution of coherences. The error can be reduced  by  further increasing $\Delta$ (or to some extent by decreasing $\Omega$) and thus extending $\tau$. 
 We note that in this (second) approach, any probe light polarization can be used.  Finally, the error can sometimes be significantly reduced by combining the two approaches. For example, adding $B = 2$T to the above example of off-resonant detection in Ca yields $p \approx 4 \times 10^{-4}$.
Depending on time constraints, available magnetic fields and laser power, as well as on the desired $N$ and $p$, the parameters can be further optimized and adiabatic switching of $\Omega$ can be considered.




To make detection spatially selective, we can apply the dark-state-based single-site addressability  \cite{gorshkov08}, shown in Fig.\ \ref{fig2}(b) on the example of Yb in the Paschen-Back regime. Let
$|r\rangle$ be the second lowest $^1S_0$ state [Fig.\ \ref{fig1}(a)].  
In addition to the probe laser $\Omega$, we apply a spatially varying control field $\Omega_c(x)$ coupling $|e\rangle$ and $|r\rangle$ in two-photon resonance with $\Omega$. 
If $\Omega_c(x)$ vanishes at the position of atom 1 and is strong on all the other atoms affected by $\Omega$, only atom 1 will fluoresce, while all other atoms will be unaffected. For example, in the first Yb example above, application of $\Omega_c/2 \pi = 1$ GHz reduces the number of scattered photons to $N \sim 0.01$ and gives only $1 \%$ decay of the $|g\rangle-|s\rangle$ coherence \cite{rnote}. 
 Alternatively, we can temporarily transfer, as described above, all Zeeman levels of $|s\rangle$ up to $|s'\rangle$ in all but one atom, apply a NOT 
 gate on all electronic qubits, carry out the detection, and then undo the NOT 
 gate and the Raman transfer. Finally, temporary lattice expansion and magnetic gradients \cite{daley} can also be used.

\textit{Conclusion.}---We have shown how to implement and couple quantum registers based on individual alkaline-earth-like atoms trapped in individual sites of an optical lattice. These quantum registers can be used as a starting point for fault-tolerant circuit-based quantum computation \cite{jiang}. Alternatively, they can be used  for high fidelity generation (and measurement) of two-colorable graph states \cite{aschauer05,  jiang}, 
which include cluster states for the use in measurement-based quantum computation \cite{raussendorf01} and GHZ states for the use in precision measurements \cite{leibfried04}. 
In particular, a cluster state can be generated in a highly parallel fashion  \cite{aschauer05} by first preparing all the electrons in state $|g\rangle + |s\rangle$ and then applying the two-qubit phase gate on each edge, which our scheme allows to do in 2 steps per each dimension of the lattice. We note that assuming high fidelity detection or a restricted error model, a four-qubit register ($I \geq 7/2$) is sufficient for the fault-tolerant operation of a quantum register \cite{jiang}. 
However, even one ($I = 1/2$) or two ($I = 3/2$) extra qubits can be used to do simpler entanglement pumping and, thus, increase the fidelity of two-colorable-graph-state generation \cite{aschauer05}.  With its accessibility using current experimental techniques and with the possibility to convert the electronic qubits into flying qubits, our approach and its extensions to ions with similar internal structure should be immediately useful in 
fields such as precision measurements, quantum computation, and quantum communication.


We thank  P.\ Julienne for discussions on scattering properties of Sr, and E.\ Demler,  I.\ Deutsch, S.\ F\"olling, L.\ Jiang, T.\ Loftus, G.\ Pupillo, I.\ Reichenbach, A.\ S{\o}rensen  for 
discussions. This work was supported by  
NSF, CUA, DARPA, AFOSR MURI, NIST, 
Austrian Science Foundation through 
SFB F40, and 
EU Network NAMEQUAM.


\vspace{-0.2in}

\begin{thebibliography}{99}

\bibitem{dur03sorensen03moehring07saffman} W.\ D\"ur and H.-J.\ Briegel, Phys.\ Rev.\ Lett.\ \textbf{90}, 067901 (2003); A.\ S.\ S{\o}rensen and K.\ M{\o}lmer, \textit{ibid.}
~\textbf{91}, 097905 (2003);
D.\ L.\ Moehring \textit{et. al.}, J. Opt. Soc. Am. B \textbf{24}, 300 (2007); M. Saffman and K. M{\o}lmer, arXiv:0805.0440v1.

\bibitem{childress05} L.\ Childress \textit{et.\ al.}, Phys.\ Rev.\ A \textbf{72}, 052330 (2005).

\bibitem{jiang} L.\ Jiang  \textit{et. al.},
e-print arXiv:quant-ph/0703029; 
Phys.\ Rev.\ A \textbf{76}, 062323 (2007).

\bibitem{dutt07jiang08stutt} M.\ V.\ G.\ Dutt \textit{et. al.}, 
Science \textbf{316}, 1312 (2007); P. Neumann \textit{et.\ al.}, \textit{ibid.}~\textbf{320}, 1326 (2008);
L.\ Jiang \textit{et.\ al.}, Phys.\ Rev.\ Lett. \textbf{100}, 073001 (2008).



\bibitem{chaudhury07} S.\ Chaudhury \textit{et. al.}, 
Phys.\ Rev.\ Lett.\ \textbf{99}, 163002 (2007).

\bibitem{ahn00} J.\ Ahn, T.\ C.\ Weinacht, and P.\ H.\ Bucksbaum, Science \textbf{287}, 463 (2000).

\bibitem{reichenbach07} I.\ Reichenbach and I.\ H.\ Deutsch, Phys.\ Rev.\ Lett.\ \textbf{99}, 123001 (2007).

\bibitem{hayes07} D.\ Hayes, P.\ S.\ Julienne, and I.\ H.\ Deutsch, Phys.\ Rev.\ Lett.\ \textbf{98}, 070501 (2007).


\bibitem{daley} A.\ J.\ Daley \textit{et. al.}, 
Phys. Rev. Lett. \textbf{101}, 170504 (2008).

\bibitem{stock08} R.\ Stock  \textit{et. al.},
Phys.\ Rev.\ A  \textbf{78}, 022301 (2008).

\bibitem{folling07} S.\ F\"olling \textit{et. al.},
Nature (London) \textbf{448}, 1029 (2007). 

\bibitem{cheinet08} P.\ Cheinet  \textit{et. al.},
Phys.\ Rev.\ Lett.\ \textbf{101}, 090404 (2008).

\bibitem{boyd07} M. M. Boyd  \textit{et. al.}, Phys. Rev. A \textbf{76}, 022510 (2007).


\bibitem{ye08} 
A. D. Ludlow \textit{et al.}, Science \textbf{319}, 1805 (2008).

\bibitem{brennen05} G.\ K.\ Brennen, D.\ P.\ O'Leary, and S.\ S.\ Bullock, Phys.\ Rev.\ A \textbf{71}, 052318 (2005).

\bibitem{speednote} If the limit on gate speed imposed by the differential $g$-factor becomes too restrictive, nuclear spin selectivity can instead be achieved with Raman transitions from $|s\rangle$ to ${}^3P_2$  via ${}^3S_1$  [Fig.\ \ref{fig1}(a)] \cite{daley}. 
One can also use Raman transitions between Zeeman levels of $|g\rangle$ via hyperfine levels of ${}^1P_1$. 

\bibitem{peil03} S.\ Peil \textit{et.\ al.}, Phys.\ Rev.\ A \textbf{67}, 051603(R) (2003).

\bibitem{cho07} J.\ Cho, Phys.\ Rev.\ Lett.\ \textbf{99}, 020502 (2007).

\bibitem{gorshkov08} A.\ V.\ Gorshkov \textit{et.\ al.}, 
Phys.\ Rev.\ Lett.\ \textbf{100}, 093005 (2008). 







\bibitem{greiner02} M.\ Greiner  \textit{et. al.},
Nature (London) \textbf{415}, 39 (2002).

\bibitem{porto} J.\ Sebby-Strabley \textit{et. al.}, Phys.\ Rev.\ A \textbf{73}, 033605 (2006).

\bibitem{lightshiftnote} We estimate that we can neglect vector and tensor light shifts, as well as the effects of hyperfine mixing \cite{boyd07} on the $g_I$-factor in $|e\rangle$. In particular, our interregister gate assumes that differential vector and tensor light shifts between the two sites are $\ll J$, which is typically satisfied.

\bibitem{roth66} L.\ M.\ Roth, Phys.\ Rev.\ \textbf{149}, 306 (1966).

\bibitem{ho98} T.-L.\ Ho, Phys.\ Rev.\ Lett.\ \textbf{81}, 742 (1998).



\bibitem{boyd07b} M.\ M.\ Boyd \textit{et. al.}, Phys.\ Rev.\ Lett.\ \textbf{98}, 083002 (2007).

\bibitem{strauch08} F.\ W.\ Strauch \textit{et. al.},
Phys.\ Rev.\ A \textbf{77}, 050304(R) (2008).



\bibitem{trotzky08} S.\ Trotzky \textit{et. al.}, Science \textbf{319}, 295 (2008).

\bibitem{lellouch87loftus02xu03} L.\ P.\ Lellouch and L.\ R.\ Hunter, Phys.\ Rev.\ A \textbf{36}, 3490 (1987);
T.\ Loftus, J.\ R.\ Bochinski, and T.\ W.\ Mossberg, \textit{ibid.}
\textbf{66}, 013411 (2002);
X.\ Xu  \textit{et. al.}, 
J.\ Opt.\ Soc.\ Am.\ B \textbf{20}, 968 (2003).

\bibitem{branchingnote} If $p$ becomes limited by the branching ratio of $|e\rangle$ decay, one should consider Be, Mg,  Zn, Cd, and Hg, which have a radiatively closed $|g\rangle-|e\rangle$ transition.


\bibitem{hoyt05} C.\ W.\ Hoyt \textit{et. al.}, Phys.\ Rev.\ Lett.\ \textbf{95}, 083003 (2005).



\bibitem{nortershauser98} W.\ N\"ortersh\"auser \textit{et. al.}, 
Spectrochim.\ Acta B \textbf{53}, 709 (1998).

\bibitem{nielsen02}  M.\ A.\ Nielsen, Phys.\ Lett.\ A \textbf{303}, 249 (2002).

020501 (2007).


\bibitem{rnote} $|r\rangle$ decay rate is $\Gamma_r = (2 \pi) 3.5$ MHz \cite{bai87}. The population of $|r\rangle$ is $\sim \Omega^2/\Omega_c^2$, so the dominant error is $\sim \tau \Gamma_r \Omega^2/\Omega_c^2$.



\bibitem{aschauer05} H.\ Aschauer, W.\ D\"ur, and H.-J.\ Briegel, Phys. Rev. A \textbf{71}, 012319 (2005).

\bibitem{raussendorf01} R.\ Raussendorf and H.-J.\ Briegel, Phys.\ Rev.\ Lett.\ \textbf{86}, 5188 (2001).

\bibitem{leibfried04} D.\ Leibfried \textit{et. al.}, Science \textbf{304}, 1476 (2004).




\bibitem{bai87} Y.\ S.\ Bai and T.\ W.\ Mossberg, Phys.\ Rev.\ A \textbf{35}, 619 (1987).







\end{thebibliography}
\end{document}